\documentclass{article}

\usepackage[preprint]{neurips_2019}

\usepackage[utf8]{inputenc} 
\usepackage[T1]{fontenc}    
\usepackage{hyperref}       
\usepackage{url}            
\usepackage{booktabs}       
\usepackage{amsfonts}       
\usepackage{nicefrac}       
\usepackage{microtype}      
\usepackage{graphicx}
\title{Use of Available Data To Inform The COVID-19 Outbreak in South Africa: A Case Study}

\author{%
  Vukosi Marivate \\
  University of Pretoria, South Africa\\
  CSIR, South Africa\\
  \texttt{vukosi.marivate@cs.up.ac.za}\\
  \And
  Herkulaas MvE Combrink \\
  University of Pretoria, South Africa\\
  University of the Free State, South Africa\\
  \texttt{CombrinkHM@ufs.ac.za}\\
}

\begin{document}

\maketitle

\begin{abstract}
The coronavirus disease (COVID-19), caused by the SARS-CoV-2 virus, was declared a pandemic by the World Health Organization (WHO) in February 2020. Currently, there are no vaccines or treatments that have been approved after clinical trials. Social distancing measures, including travel bans, school closure, and quarantine applied to countries or regions are being used to limit the spread of the disease, and the demand on the healthcare infrastructure. The seclusion of groups and individuals has led to limited access to accurate information. To update the public, especially in South Africa, announcements are made by the minister of health daily. These announcements narrate the confirmed COVID-19 cases and include the age, gender, and travel history of people who have tested positive for the disease. Additionally, the South African National Institute for Communicable Diseases updates a daily infographic summarising the number of tests performed, confirmed cases, mortality rate, and the regions affected. However, the age of the patient and other nuanced data regarding the transmission is only shared in the daily announcements and not on the updated infographic. To disseminate this information, the Data Science for Social Impact research group at the University of Pretoria, South Africa, has worked on curating and applying publicly available data in a way that is computer readable so that information can be shared to the public - using both a data repository and a dashboard.. Through collaborative practices, a variety of challenges related to publicly available data in South Africa came to the fore. These include shortcomings in the accessibility, integrity, and data management practices between governmental departments and the South African public. In this paper, solutions to these problems will be shared by using a publicly available data repository and dashboard as a case study.

Dashboard: \url{https://bitly.com/covid19za-dash}\\
Data repository: \url{https://github.com/dsfsi/covid19za}
\end{abstract}



\section{Introduction}
Accurate data are at the centre of mitigating risk, preventing widespread panic and sensationalism during a natural disaster. Evidence-based information obtained from accurate data is an asset, and one of the only strategic resources the public have during a crisis.  The practice of sharing information to the public about the current state of things is dependent on specific data that has to be captured and shared to the public in a way that is useful, usable and desirable.  During a crisis, the public needs to minimize exposure to the situation or act accordingly to provide support where needed. The contribution of this paper is a framework that substantiates the type of data employed to capture and modify shared information with the public during a crisis of a biological nature, such as the COVID-19 pandemic. This paper summarises findings (e.g., demography and neighbourhood) based on the public data repository~\cite{marivate_vukosi_2020_3732419} and dashboard\footnote{\url{https://bitly.com/covid19-dash}}, to support general understanding and lessons learned from the COVID-19 epidemic.

The disease, COVID-19, is a severe acute respiratory syndrome (SARS) caused by the SARS-CoV-2 virus \cite{Bai01}. The first reported case of COVID-19 was in December 2019 in Wuhan, China \cite{Phingzeng10}. The problem with SARS-CoV-2 is the pressure it puts on the healthcare system because of its high infection rate \cite{Zhao12}. Globally, sudden spikes in the confirmed COVID-19 cases are severe because there are limited resources to effectively manage and treat patients in any of the current healthcare systems \cite{Ferguson04}. As of February 2020, the World Health Organization (WHO) declared SARS-CoV-2 a pandemic- one of the first biological threats to modern society in the 21st century.\cite{Remuzzi09}. In order for the South African government to reduce the widespread of SARS-CoV-2, travelling restrictions were placed on high-risk regions and border-crossings were closed from one country to another. South Africa experienced the first COVID-19 case on the 28th of February 2020, and as of the 25th of March 2020, the number of confirmed COVID-19 cases increased to 702, despite efforts to contain the virus by putting a ban on international travel \cite{WHO11}. The increase in confirmed cases may cause widespread panic and anxiety,  which is why the public relies on good, reliable information and data, now more than ever. 

Currently, information regarding the COVID-19 outbreak in South Africa is shared with the public across various platforms, of which two are most popular/most widely used. Firstly, the National Institute for Communicable Diseases (NICD) publishes an infographic that contains limited information, providing a bird's-eye view of the outbreak. This information is limited to South Africa and only reports the number of tests performed, number of confirmed cases, which regions are affected, and COVID-19 related mortality \cite{NICD08}. Secondly, the minister of the Department of Health (DoH) in South Africa, Zwelini Mkhize, updates the public regarding the cases as they arise on a daily basis. These updates are published sentences on the DoH website, containing some demographic information about the confirmed cases, including age, gender, travel history and mode of contraction of SARS-CoV-2 \cite{DoH03}. Although these sources of information are valuable, they are potentially ineffective ways of sharing information to the public regarding their usability for a variety of reasons. Amongst these is the number of different platforms a person has to navigate through to gain access to accurate data. Additionally, the format in which the data are presented is not in a computer readable format and has to undergo processing in order to be used and stored. This further complicates legibility, simplicity and accessibility of the information that is shared, a concern about South African government data that was highlighted in prior work~\cite{marivate2018exploring}. The impact of not having useful, usable and desirable information has a direct effect on management strategies and responses from the public in relation to the disease \cite{Gonulal05}. 

\begin{figure}
\includegraphics[width=0.75\columnwidth]{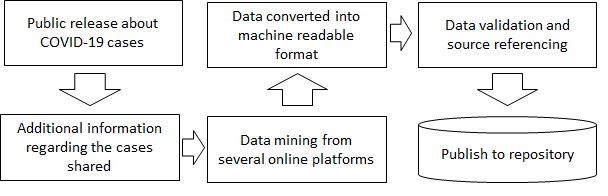}
\caption{Data publishing cycle of COVID-19 data}
\label{fig:data_cycle}
\end{figure}

To counter the aforementioned problem, the Data Science for Social Impact (DSFSI) research group at the University of Pretoria, South Africa, has developed an open repository for the data integrity of South African COVID-19 cases. DSFSI Lab members and willing volunteers are responsible for the mining, validation and storage of data related to the COVID-19 patients in South Africa. To work collectively on a project related to public data in a way that can be scaled, the DSFSI manages and consolidates the available data related to the COVID-19 cases in South Africa. Once consolidated, the data are shared in an open, publicly available repository on GitHub.com, and then linked to the dashboard \cite{Marivate07}. On the repository, any member or user has the freedom to critique and propose new features or data to be added to the repository or dashboard. This includes data integrity issues as well as information to be added to the repository that is otherwise outdated or virtually inaccessible to the public. The workflow of this process is illustrated in Fig. \ref{fig:data_cycle}.

In order to link information to the COVID-19 records, data needs to be accurate and restructured. Unfortunately, when data are not updated and information changes about the state of the evidence, then it becomes difficult to track when the changes occurred, especially if the changes relate to the unique ID or name of the item. One such example relates to the publicly available South African hospital data. Some items in the hospital data were last updated more than two years prior to 2020. This includes hospital data such as coordinates, accurate contact information, and facilities available within the hospital, as well as the population size of a/the district. 

Most of the aforementioned information is not consolidated in one place a great risk factor in a time where the public needs to know where to seek care, and which hospitals are equipped to test and manage COVID-19 patients. Another example came about in a sudden spike of confirmed cases between the 23rd and 25th of March 2020.  This increase of 435 cases came unaccompanied by any of the aforementioned demographic information, transmission type, nor their travel history, and these results are still pending. In subsequent days, there has been the same challenge, although the latest available data on deaths has demographic information and some travel history.

\section{Aims and Methodology}

The primary objectives of this study were to determine what data should be included in a public repository amidst the COVID-19 outbreak and how this data should be disseminated within a public dashboard. The public repository of data followed a Creative Commons licence for data, and MIT License for Code, with copyright for the Data Science for Social Impact research group at the University of Pretoria, South Africa \cite{jsoup06}.  All of the data were gathered and consolidated on the public repository which is hosted on GitHub\footnote{\url{https://www.github.com/dsfsi/covid19za}} and uploaded on Zenodo~\cite{marivate_vukosi_2020_3732419}.

\begin{figure}[h]
\includegraphics[width=0.49\columnwidth]{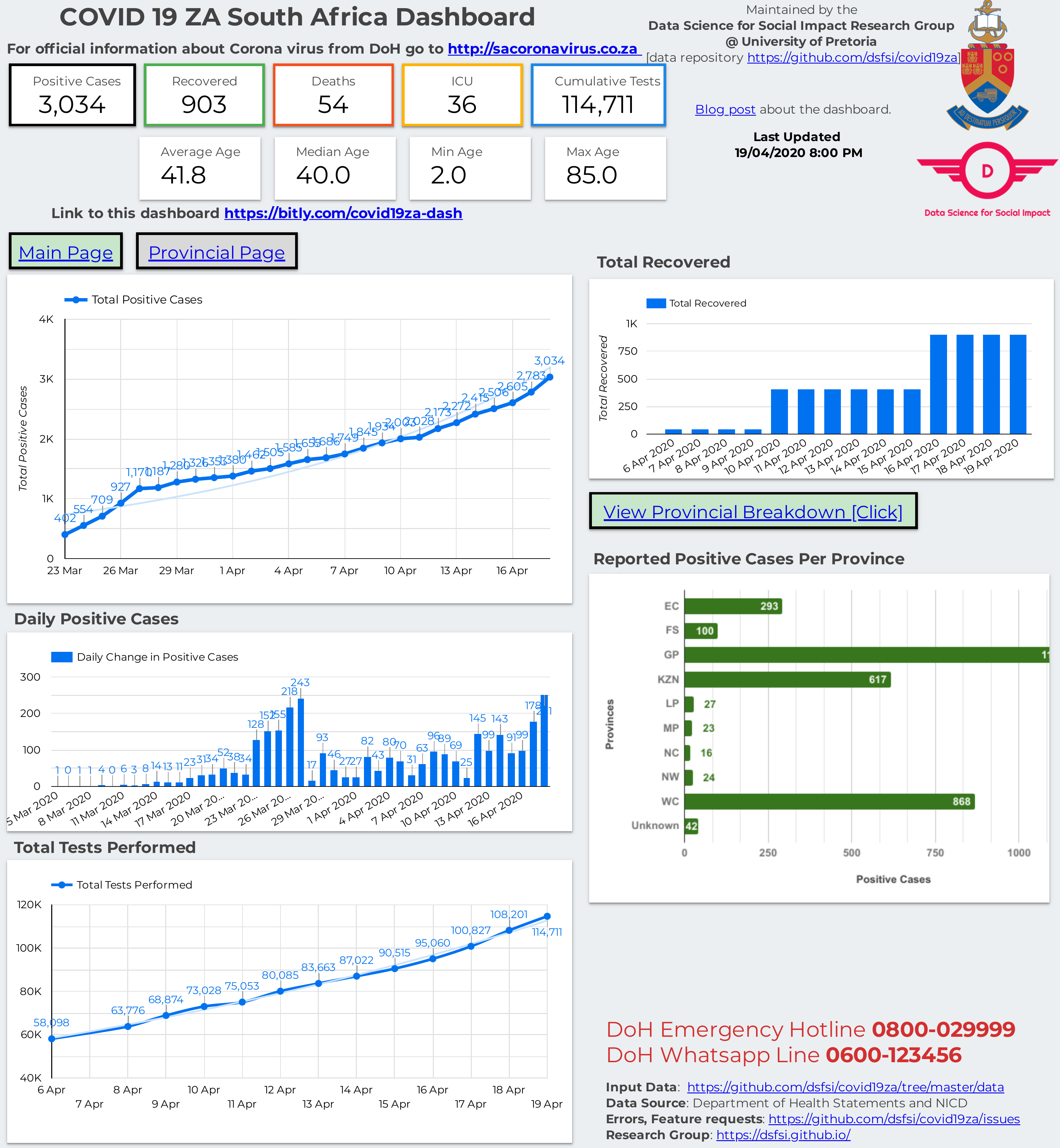}
\includegraphics[width=0.49\columnwidth]{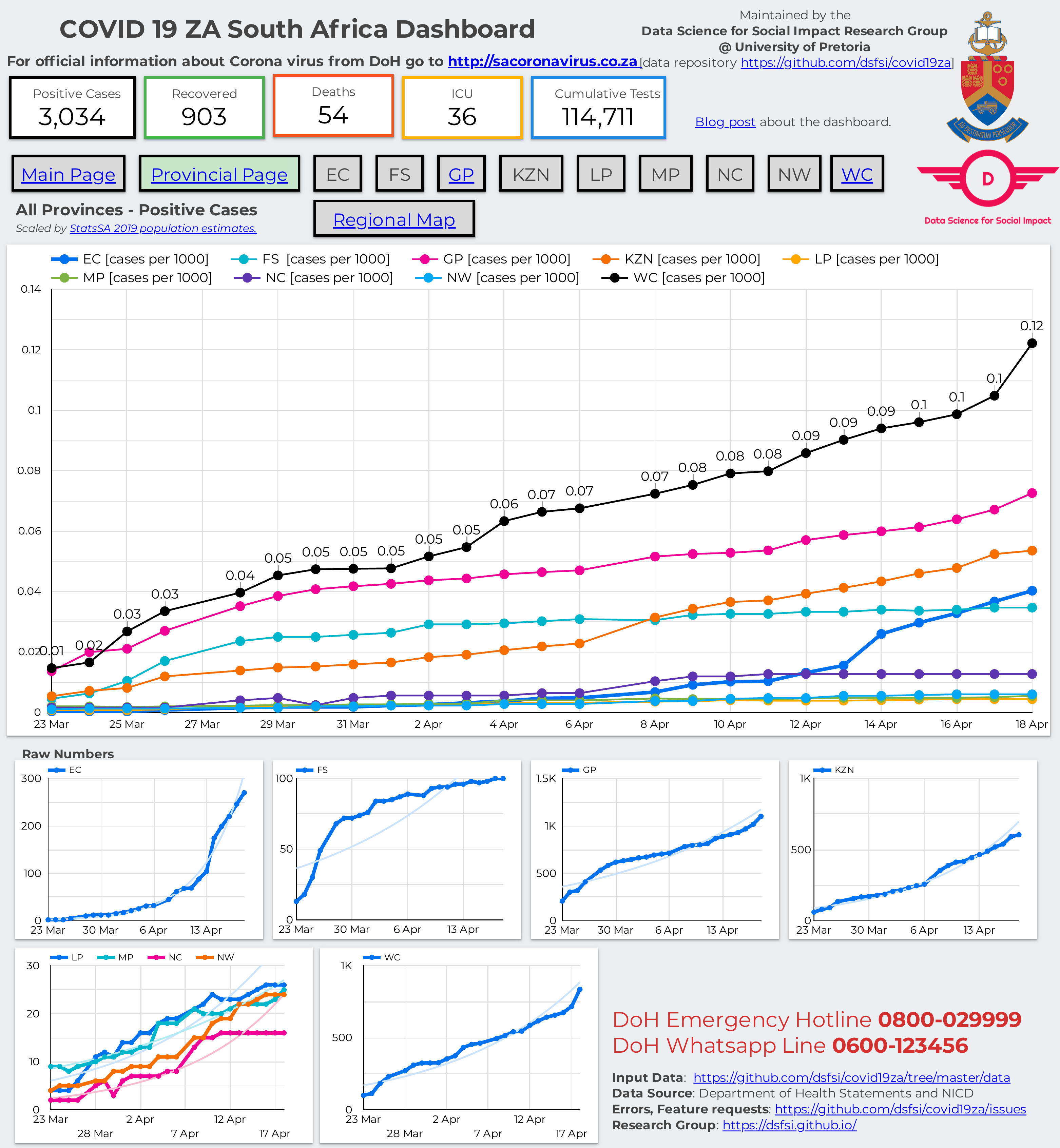}
\caption{Consolidated dashboard using data from the repository. \emph{Left}: Front page with aggregated national statistics. \emph{Right}: Aggregated statistics by province.}
\label{fig:consolidated_dashboard}
\end{figure}

To determine if the dashboard (Fig. \ref{fig:consolidated_dashboard}) and data repository were used, data and analytics were performed on the basis of descriptive statistics related to the number of views, clicks, comments and recommendations on the public repository and dashboard. The use of the repository goes further than just the views, but that other researchers actually use it for analysis.

To measure whether or not the repository and dashboard were useful and desirable, the public repository allowed for the posting of issues, comments and recommendations. These items were categorised according to their submission on the repository. An item could be sorted into more than one category depending on the nature of the problem. The DSFSI and public have the opportunity to choose which problems they would like to work on, and solutions are approved by the DSFSI group.

\section{Results}
\begin{figure}[h]
\includegraphics[width=0.75\columnwidth]{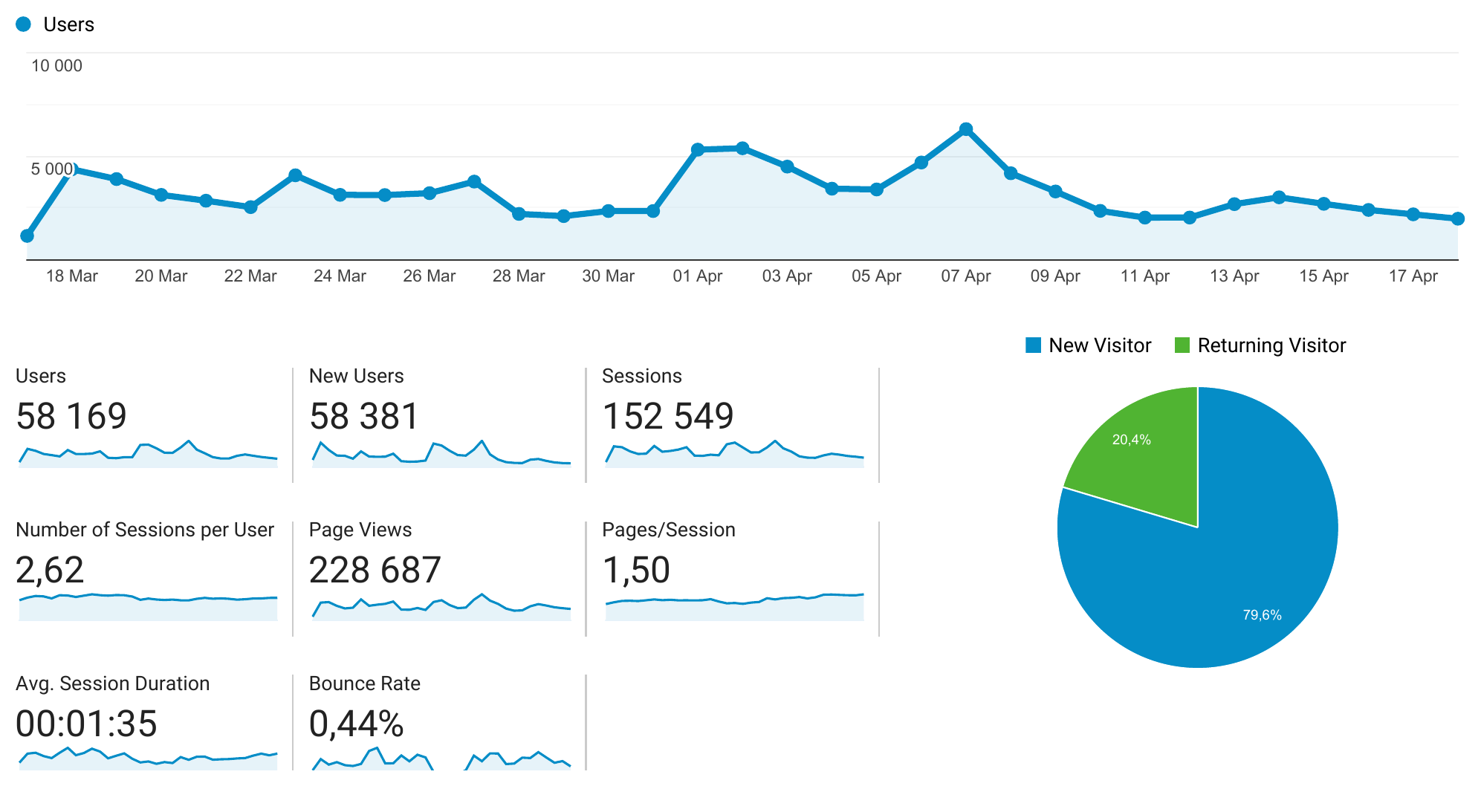}
\caption{Data usage information from the website.}
\label{fig:data_usage}
\end{figure}

In total, 58,169 users accessed the dashboard in the time period of 17th March to 18th April 2020 (Fig. \ref{fig:data_usage})

As seen in Fig. \ref{fig:data_usage}, at least 20.4 percent of users return to the dashboard, with an average session duration of 1.35 minutes per session. In addition to this, the repository from which the data are drawn have had over 2,000 clones of the dataset from at least 200 different people. The majority of the users were from the GitHub community, but a few unique users (meaning they had not visited before) accessed the repository from other platforms (Table \ref{tab:data_usage}).

\begin{table}[h]
  \caption{Data usage information from the website}
  \label{tab:data_usage}
  \begin{tabular}{ccl}
    \toprule
    Non-English or Site&Views&Unique visitors\\
    \midrule
    \ github.com & 2,229 & 797\\
    \ twitter.com & 675 & 431\\
    \ Google.com & 647 & 353\\
    \ Linkedin.com & 212 & 125\\
    \ zendo.org & 58 & 26\\
    \ Bing.com & 57 & 27\\
    \ facebook.com & 43 & 27\\
    \ m.facebook.com & 39 & 26\\
    \ vima.co.za & 20 & 9\\
    \ Ink.in & 17 & 8\\
  \bottomrule
\end{tabular}
\end{table}

At least 404 unique visitors viewed the issues that were posted on the repository. A further 150 unique visitors viewed the pull requests. These were requests for data that members of the DSFSI and the public contributed to the actual datasets. To manage the issues that were posted, ten different labels were created to categorise the issues. Currently, only six of the ten labels were used for the issues that were posted. These were: bugs, data, enhancements, good first issue, help wanted and questions (Table \ref{tab:issue_categories}). 

\begin{table}[h]
  \caption{Categories of lodged issues}
  \label{tab:issue_categories}
  \begin{tabular}{ccl}
    \toprule
    Label& Resolved issues & Unresolved issues \\
    \midrule
    \ Bugs &  7 & 2 \\
    \ Data & 19 & 10\\
    \ Enhancement & 26 & 17\\
    \ Good first issue & 2 & 2\\
    \ Help wanted & 2 & 1\\
    \ Question & 1 & 1\\
      \bottomrule
\end{tabular}
\end{table}

Bugs referred to any error in the data or issue related to a feature of either the dashboard or function within the repository. The single unresolved issue in bugs related to a single incorrect data entry, but finding the source confirming the correct data proved challenging. Data were any inquiry about the data including differences between data sets, missing data, or additions to datasets. To resolve these issues, data needs to be updated from the source. Enhancement meant improvements to current implementations of either the data in the repository or dashboard. Pending information for most of these enhancements. These include additional fields that were not provided in the publicly available data. Good first issues were entry level problems that could be completed by people from any background. These were labelled as such for newcomers to the project that did not require either a lot of time or expertise to work on. Help wanted translated to problems that require additional attention. The unresolved issues require data that are not currently available to the public in order to solve the issues. Questions were presented as general request for clarity or required more information on a particular issue in the repository that another person posted. To resolve the one unresolved question, a decision has to be made internally about the data to resolve the matter. There were more than 10,000 additions to the repository data, and 1,430 deletions, all reviewed by different members of DSFSI and accepted if they were noteworthy contributions to the repository. In addition to this, there were 26 different contributors who pushed 345 commits to all branches within the repository.

\begin{figure}[h]
\includegraphics[width=0.49\columnwidth]{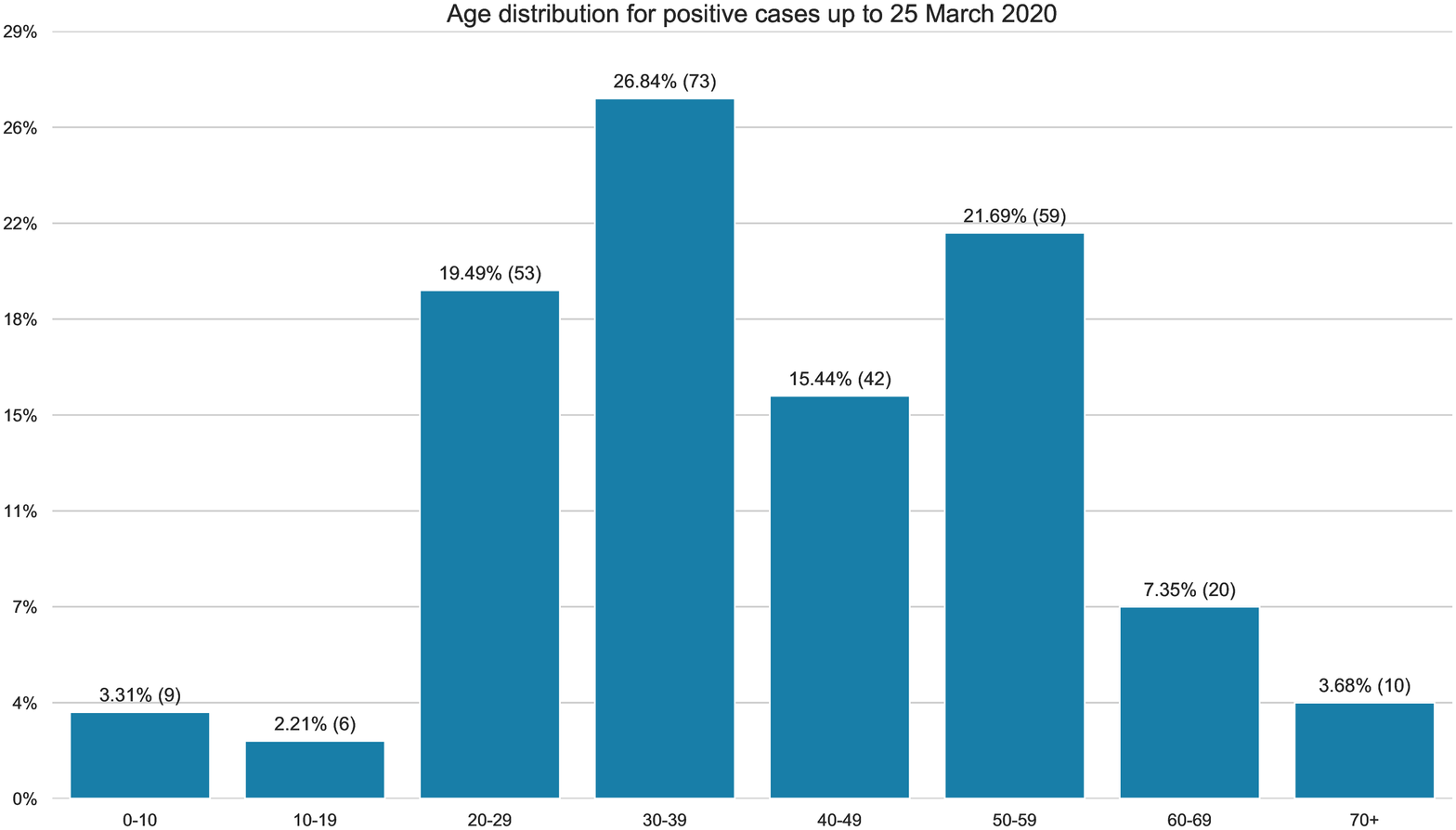}
\includegraphics[width=0.49\columnwidth]{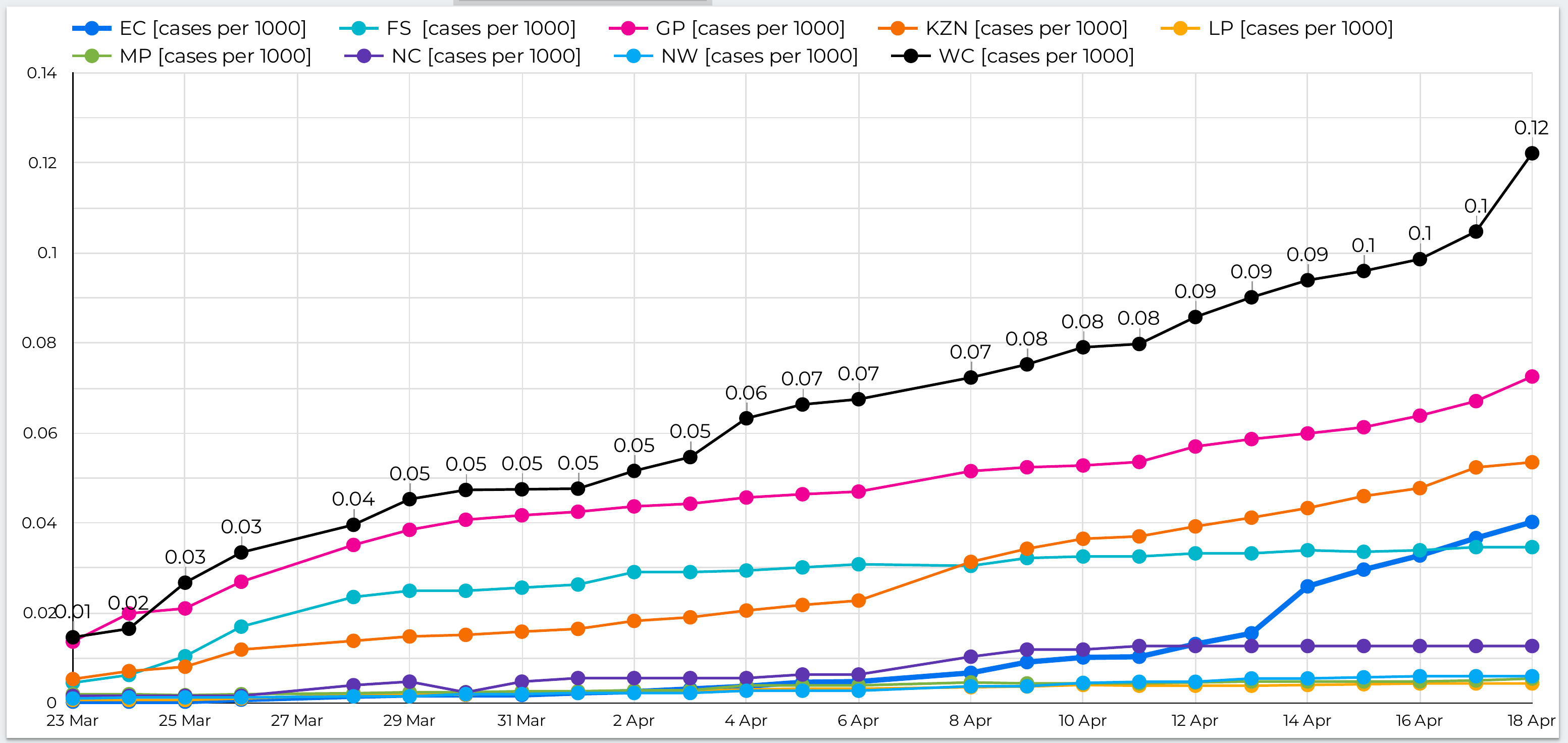}
\caption{Examples of analysis made possible by the data repository. \emph{Left} Age distribution of positive cases. \emph{Right:} Provincial growth}
\label{fig:analysis}
\end{figure}

The majority of requests and changes to the repository and dashboard were associated with the data or enhancements to the data. In total, there are fifteen datasets, with seven of them related to information about hospitals in SA.  Once created, the subsequent data were displayed in a dashboard \cite{marivate_vukosi_2020_3732419}. Included in the dashboard were information related to COVID-19, a South African helpline, sources of the information, when last the information was updated, a blog post containing the purpose of the dashboard, links to the open public repository, and general information about the research group. Some analysis that can be accomplished with the data is shown in Fig. \ref{fig:analysis}. 
 
\section{Conclusion}
Data are one of the most important assets during a crisis. Unfortunately, not prioritising this commodity had complications during the early days of the COVID-19 pandemic from a South African perspective. To prevent this from happening, the DSFSI research group have started collaborating and expanding this type of methodology to create a line list for the rest of the African continent\footnote{\url{https://www.github.com/dsfsi/covid19africa}}. The data from this project led to discussion between DSFSI and the NICD and DoH, in an attempt to assist the situation. The COVID-19 pandemic showed the world the value of data, and that people and systems should effectively prepare for a time of crisis. Enabling the public to engage with data in a way that is open and collaborative is an overlooked service that can aid during a calamity. Furthermore, having readily available data is useful when needed during an emergency, but can seem redundant during peaceful times. Therefore, prioritising data management practices, getting input from the people who use the information, and collaborating between different organisations to gain the same result should be a proactive approach and standard, not one that is only implemented during a catastrophe.  

\section{Acknowledgement}
We would like to acknowledge every person from the public that dedicated their time, effort and energy to assist during this pandemic. At the time of this publication, in alphabetical order, Alta de Waal, Jay Welsh, Nompumelelo Mtsweni, Ofentswe Lebogo, Shiven Moodley, Vutlhari Rikhotso, S'busiso Mkhondwane. As this is an open contribution project, the updated list of contributors is available on the github repo\footnote{\url{https://github.com/dsfsi/covid19za/graphs/contributors}}. We would like to thank the DSFSI research group at the University of Pretoria for all their expertise, patience and hard work during this time. We also would like to thank all the employees of the NICD, DoH and WHO who assisted with data during this time.  We would like to acknowledge ABSA for sponsoring the industry chair and it's related activities to the project.

\bibliographystyle{unsrt}
\bibliography{reference_coviddsjournal}


\begin{thebibliography}{13}


\ifx \showCODEN    \undefined \def \showCODEN     #1{\unskip}     \fi
\ifx \showDOI      \undefined \def \showDOI       #1{#1}\fi
\ifx \showISBNx    \undefined \def \showISBNx     #1{\unskip}     \fi
\ifx \showISBNxiii \undefined \def \showISBNxiii  #1{\unskip}     \fi
\ifx \showISSN     \undefined \def \showISSN      #1{\unskip}     \fi
\ifx \showLCCN     \undefined \def \showLCCN      #1{\unskip}     \fi
\ifx \shownote     \undefined \def \shownote      #1{#1}          \fi
\ifx \showarticletitle \undefined \def \showarticletitle #1{#1}   \fi
\ifx \showURL      \undefined \def \showURL       {\relax}        \fi
\providecommand\bibfield[2]{#2}
\providecommand\bibinfo[2]{#2}
\providecommand\natexlab[1]{#1}
\providecommand\showeprint[2][]{arXiv:#2}

\bibitem[\protect\citeauthoryear{Bai, Yao, and Wei}{Bai et~al\mbox{.}}{2020}]%
        {Bai01}
\bibfield{author}{\bibinfo{person}{Yan Bai}, \bibinfo{person}{Lingsheng Yao},
  {and} \bibinfo{person}{Tao Wei}.} \bibinfo{year}{2020}\natexlab{}.
\newblock \showarticletitle{Presumed asymptomatic carrier transmission of
  COVID-19}.
\newblock \bibinfo{journal}{\emph{Jama}} (\bibinfo{date}{March}
  \bibinfo{year}{2020}), \bibinfo{pages}{E1}.
\newblock
\urldef\tempurl%
\url{https://doi.org/10.1001/jama.2020.2565}
\showDOI{\tempurl}


\bibitem[\protect\citeauthoryear{DoH}{DoH}{2020}]%
        {DoH03}
\bibfield{author}{\bibinfo{person}{DoH}.} \bibinfo{year}{2020}\natexlab{}.
\newblock \bibinfo{booktitle}{\emph{South African Department of Health}}.
\newblock
\urldef\tempurl%
\url{http://www.health.gov.za/}
\showURL{%
Retrieved March 26, 2020 from \tempurl}


\bibitem[\protect\citeauthoryear{Ferguson, Laydon, Nedjati-Gilani, Imai,
  Ainslie, Baguelin, Bhatia, Boonyasiri, CucunubÃ¡, Cuomo-Dannenburg, Dighe,
  Dorigatti, Fu, Gaythorpe, Green, Hamlet, Hinsley, Okell, van Elsland,
  Thompson, Verity, Volz, Wang, Wang, Walker, Winskill, Whittaker, Donnelly,
  Riley, and Ghani}{Ferguson et~al\mbox{.}}{2020}]%
        {Ferguson04}
\bibfield{author}{\bibinfo{person}{Neil~M Ferguson}, \bibinfo{person}{Daniel
  Laydon}, \bibinfo{person}{Gemma Nedjati-Gilani}, \bibinfo{person}{Natsuko
  Imai}, \bibinfo{person}{Kylie Ainslie}, \bibinfo{person}{Marc Baguelin},
  \bibinfo{person}{Sangeeta Bhatia}, \bibinfo{person}{Adhiratha Boonyasiri},
  \bibinfo{person}{Zulma CucunubÃ¡}, \bibinfo{person}{Gina Cuomo-Dannenburg},
  \bibinfo{person}{Amy Dighe}, \bibinfo{person}{Ilaria Dorigatti},
  \bibinfo{person}{Han Fu}, \bibinfo{person}{Katy Gaythorpe},
  \bibinfo{person}{Will Green}, \bibinfo{person}{Arran Hamlet},
  \bibinfo{person}{Wes Hinsley}, \bibinfo{person}{Lucy~C Okell},
  \bibinfo{person}{Sabine van Elsland}, \bibinfo{person}{Hayley Thompson},
  \bibinfo{person}{Robert Verity}, \bibinfo{person}{Erik Volz},
  \bibinfo{person}{Haowei Wang}, \bibinfo{person}{Yuanrong Wang},
  \bibinfo{person}{Patrick~GT Walker}, \bibinfo{person}{Peter Winskill},
  \bibinfo{person}{Charles Whittaker}, \bibinfo{person}{Christl~A Donnelly},
  \bibinfo{person}{Steven Riley}, {and} \bibinfo{person}{Azra~C Ghani}.}
  \bibinfo{year}{2020}\natexlab{}.
\newblock \showarticletitle{Impact of non-pharmaceutical interventions (NPIs)
  to reduce COVID19 mortality and healthcare demand}.
\newblock \bibinfo{journal}{\emph{London: Imperial College COVID-19 Response
  Team}} (\bibinfo{date}{March} \bibinfo{year}{2020}), \bibinfo{pages}{1 --
  20}.
\newblock
\urldef\tempurl%
\url{https://doi.org/10.25561/77482}
\showDOI{\tempurl}


\bibitem[\protect\citeauthoryear{Gonulal}{Gonulal}{2019}]%
        {Gonulal05}
\bibfield{author}{\bibinfo{person}{Talip Gonulal}.}
  \bibinfo{year}{2019}\natexlab{}.
\newblock \showarticletitle{Missing data management practices in L2 research:
  the good, the bad and the ugly}.
\newblock \bibinfo{journal}{\emph{Erzincan University Education Faculty
  Journal}} \bibinfo{volume}{1}, \bibinfo{number}{21} (\bibinfo{date}{Feb.}
  \bibinfo{year}{2019}), \bibinfo{pages}{56 -- 73}.
\newblock
\urldef\tempurl%
\url{https://doi.org/10.17556/erziefd.448559}
\showDOI{\tempurl}


\bibitem[\protect\citeauthoryear{JSOUP}{JSOUP}{2020}]%
        {jsoup06}
\bibfield{author}{\bibinfo{person}{JSOUP}.} \bibinfo{year}{2020}\natexlab{}.
\newblock \bibinfo{booktitle}{\emph{KSOUP Licence}}.
\newblock
\urldef\tempurl%
\url{https://jsoup.org/license}
\showURL{%
Retrieved February 29, 2020 from \tempurl}


\bibitem[\protect\citeauthoryear{Marivate, de~Waal, Combrink, Lebogo, Moodley,
  Mtsweni, Rikhotso, Welsh, and Mkhondwane}{Marivate et~al\mbox{.}}{2020}]%
        {marivate_vukosi_2020_3732419}
\bibfield{author}{\bibinfo{person}{Vukosi Marivate}, \bibinfo{person}{Alta de
  Waal}, \bibinfo{person}{Herkulaas Combrink}, \bibinfo{person}{Ofentswe
  Lebogo}, \bibinfo{person}{Shivan Moodley}, \bibinfo{person}{Nompumelelo
  Mtsweni}, \bibinfo{person}{Vuthlari Rikhotso}, \bibinfo{person}{Jay Welsh},
  {and} \bibinfo{person}{S'busiso Mkhondwane}.}
  \bibinfo{year}{2020}\natexlab{}.
\newblock \bibinfo{booktitle}{\emph{{Coronavirus disease (COVID-19) case data -
  South Africa}}}.
\newblock
\urldef\tempurl%
\url{https://doi.org/10.5281/zenodo.3732419}
\showDOI{\tempurl}


\bibitem[\protect\citeauthoryear{Marivate and Group}{Marivate and
  Group}{2020}]%
        {Marivate07}
\bibfield{author}{\bibinfo{person}{Vukosi Marivate} {and} \bibinfo{person}{Data
  Science For Social Impact~Research Group}.} \bibinfo{year}{2020}\natexlab{}.
\newblock \bibinfo{booktitle}{\emph{COVID 19 ZA South Africa Dashboard}}.
\newblock
\urldef\tempurl%
\url{https://datastudio.google.com/u/0/reporting/1b60bdc7-bec7-44c9-ba29-be0e043d8534/page/hrUIB}
\showURL{%
\tempurl}
\newblock
\shownote{Accessed at
  https://datastudio.google.com/u/0/reporting/1b60bdc7-bec7-44c9-ba29-be0e043d8534/page/hrUIB.}


\bibitem[\protect\citeauthoryear{Marivate and Moorosi}{Marivate and
  Moorosi}{2018}]%
        {marivate2018exploring}
\bibfield{author}{\bibinfo{person}{Vukosi Marivate} {and}
  \bibinfo{person}{Nyalleng Moorosi}.} \bibinfo{year}{2018}\natexlab{}.
\newblock \showarticletitle{Exploring data science for public good in South
  Africa: evaluating factors that lead to success}. In
  \bibinfo{booktitle}{\emph{Proceedings of the 19th Annual International
  Conference on Digital Government Research: Governance in the Data Age}}.
  \bibinfo{pages}{1--6}.
\newblock


\bibitem[\protect\citeauthoryear{NICD}{NICD}{2020}]%
        {NICD08}
\bibfield{author}{\bibinfo{person}{NICD}.} \bibinfo{year}{2020}\natexlab{}.
\newblock \bibinfo{booktitle}{\emph{National Institute For Communicable
  Diseases (NICD)}}.
\newblock
\urldef\tempurl%
\url{http://www.nicd.ac.za/}
\showURL{%
Retrieved March 26, 2020 from \tempurl}


\bibitem[\protect\citeauthoryear{Organization}{Organization}{2020}]%
        {WHO11}
\bibfield{author}{\bibinfo{person}{World~Health Organization}.}
  \bibinfo{year}{2020}\natexlab{}.
\newblock \bibinfo{booktitle}{\emph{Coronavirus disease 2019 (COVID-19):
  situation report}}.
\newblock
\urldef\tempurl%
\url{https://apps.who.int/iris/bitstream/handle/10665/331475/nCoVsitrep11Mar2020-eng.pdf}
\showURL{%
Retrieved March 16, 2020 from \tempurl}


\bibitem[\protect\citeauthoryear{Remuzzi and Remuzzi}{Remuzzi and
  Remuzzi}{2020}]%
        {Remuzzi09}
\bibfield{author}{\bibinfo{person}{Andrea Remuzzi} {and}
  \bibinfo{person}{Giuseppe Remuzzi}.} \bibinfo{year}{2020}\natexlab{}.
\newblock \showarticletitle{COVID-19 and Italy: what next?}
\newblock \bibinfo{journal}{\emph{The Lancet}} \bibinfo{volume}{395},
  \bibinfo{number}{10229} (\bibinfo{date}{March} \bibinfo{year}{2020}),
  \bibinfo{pages}{1 -- 4}.
\newblock
\urldef\tempurl%
\url{https://doi.org/10.1016/S0140-6736(20)30627-9}
\showDOI{\tempurl}


\bibitem[\protect\citeauthoryear{Robson}{Robson}{2020}]%
        {Robson10}
\bibfield{author}{\bibinfo{person}{Barry Robson}.}
  \bibinfo{year}{2020}\natexlab{}.
\newblock \showarticletitle{COVID-19 and Italy: what next?}
\newblock \bibinfo{journal}{\emph{The Lancet}} \bibinfo{volume}{119},
  \bibinfo{number}{103670} (\bibinfo{date}{April} \bibinfo{year}{2020}),
  \bibinfo{pages}{1 -- 19}.
\newblock
\urldef\tempurl%
\url{https://doi.org/doi.org/10.1016/j.compbiomed.2020.103670}
\showDOI{\tempurl}


\bibitem[\protect\citeauthoryear{Zhao, Yu, Zha, Wang, Pang, Li, and Li}{Zhao
  et~al\mbox{.}}{2020}]%
        {Zhao12}
\bibfield{author}{\bibinfo{person}{Wen Zhao}, \bibinfo{person}{Shikai Yu},
  \bibinfo{person}{Xiangyi Zha}, \bibinfo{person}{Ning Wang},
  \bibinfo{person}{Qiumei Pang}, \bibinfo{person}{Tongzeng Li}, {and}
  \bibinfo{person}{Aixin Li}.} \bibinfo{year}{2020}\natexlab{}.
\newblock \showarticletitle{linical characteristics and durations of
  hospitalized patients with COVID-19 in Beijing: a retrospective cohort
  study}.
\newblock \bibinfo{journal}{\emph{MedRxiv}} \bibinfo{volume}{119},
  \bibinfo{number}{103670} (\bibinfo{date}{March} \bibinfo{year}{2020}),
  \bibinfo{pages}{1 -- 6}.
\newblock
\urldef\tempurl%
\url{https://doi.org/10.1101/2020.03.13.20035436}
\showDOI{\tempurl}


\end{thebibliography}

\end{document}